\documentclass{PoS}

\PoS{PoS(HEP2005)202}

\title{$D^{+} \to \mu^{+}\nu$ and $f_{D^+}$ from 281 pb$^{-1}$ 
       at $\psi$(3770) from CLEO-c}

\ShortTitle{Leptonic $D^{+}$ Decays from CLEO-c}

\author{\speaker{Yongsheng Gao (for CLEO Collaboration)}\\

        Southern Methodist University, Dallas, Texas 75275-0175, USA\\

        E-mail: \email{gao@mail.physics.smu.edu}}

\abstract{
We report improved measurements of branching fraction and 
decay constant $f_{D^+}$ in $D^{+} \to \mu^{+}\nu$ using
281 pb$^{-1}$ of data taken on the $\psi(3770)$ resonance 
with the CLEO-c detector.
We extract a relatively precise value for the decay constant of
the $D^+$ meson by measuring ${\cal{B}}(D^+\to\mu^+\nu)=(4.40\pm
0.66^{+0.09}_{-0.12})\times 10^{-4}$ and find 
$f_{D^+}=(222.6\pm 16.7^{+2.8}_{-3.4})~{\rm MeV}$. 
We also set a 90\% confidence upper limit on 
${\cal{B}}(D^+\to e^+\nu)<2.4\times
10^{-5}$ which limits contributions from non-standard model
physics.
}

\FullConference{International Europhysics Conference on High Energy Physics\\

                 July 21st - 27th 2005\\

                 Lisboa, Portugal}

\newcommand{\etal}{{\it et al.}}

\begin{document}

\section{Introduction}
The last 3 decades have witnessed amazing progress in Heavy Flavor physics.
To test the SM and search for new physics, the precise measurements of CKM 
matrix elements have been one of the focus of current HEP efforts.
However, our discovery potential is limited by our ability to relate the 
world of hadrons to the world of quarks, that is, the systematic errors due 
to non-pertubative QCD.

The CLEO-c program at $\psi$(3770) is an important part of the global efforts 
in heavy flavor physics. The precise measurements of decay constants $f_{D^+}$  
and $f_{D_{s}}$ will test the Lattice QCD calculation and gain confidence in the 
theoretical prediction of $f_{B}$.
The decay $D^+\to \ell^+\nu$ proceeds by the $c$ and $\overline{d}$
quarks annihilating into a virtual $W^+$, with a decay width
\cite{Formula1}
\begin{equation}
\Gamma(D^+\to \ell^+\nu) = {{G_F^2}\over 8\pi}f_{D^+}^2m_{\ell}^2M_{D^+}
\left(1-{m_{\ell}^2\over M_{D^+}^2}\right)^2 \left|V_{cd}\right|^2~~~,
\label{eq:equ_rate}
\end{equation}
where $M_{D^+}$ is the $D^+$ mass, $m_{\ell}$ is the mass of the
final state lepton, $|V_{cd}|$ is a CKM matrix element that we
assume to be equal to $|V_{us}|$, and $G_F$ is the Fermi coupling
constant.

\section{Data Sample and Event Selection}
In this study \cite{CLNS05-1932} we use 281 pb$^{-1}$ of data produced 
in $e^+e^-$
collisions using the Cornell Electron Storage Ring (CESR) and
recorded at the $\psi''$ resonance (3.770 GeV). 
Our analysis strategy is to fully reconstruct the $D^-$ meson in one
of six decay modes listed in Table~\ref{tab:Dreconnew} and search
for a $D^+\to \mu^+\nu$ decay in the rest of the event. 
Track selection, particle identification (PID), $\pi^0$, $K_S$, and 
muon selection cuts are identical to those used in Ref.
\cite{CLEODptomunu}.

Table~\ref{tab:Dreconnew} gives the numbers of signal and background
events for each mode within the signal region, defined as $m_D -
2.5\: \sigma_{m_{BC}}  < m_{BC} < m_D + 2.0\: \sigma_{m_{BC}}$,
where $\sigma_{m_{BC}}$ is the r.m.s. width of the lower side of the
distribution.

\begin{table}[htb]
\begin{center}
\begin{tabular}{lrcrc}\hline\hline
    Mode  &  \multicolumn{3}{c}{Signal}           &  Background \\ \hline
$K^+\pi^-\pi^- $ & 77387& $\pm$& 281   & $~1868$\\
$K^+\pi^-\pi^- \pi^0$ & 24850 &$\pm$& 214  & $12825$\\
$K_S\pi^-$ &   11162&$\pm$& 136& ~~~514\\
$K_S\pi^-\pi^-\pi^+ $ &  18176 &$\pm$& 255 & $~8976$\\
$K_S\pi^-\pi^0 $ &  20244&$\pm$& 170 & ~5223\\
$K^+K^-\pi^-$ & 6535&$\pm$& 95 &~1271 \\\hline
Sum &   158354&$\pm$& 496 & 30677\\
\hline\hline
\end{tabular}
\end{center}
\caption{Tagging modes and numbers of signal and background
events.
\label{tab:Dreconnew}}
\end{table}

\begin{figure}
{\mbox{\includegraphics[width=0.50\textwidth]{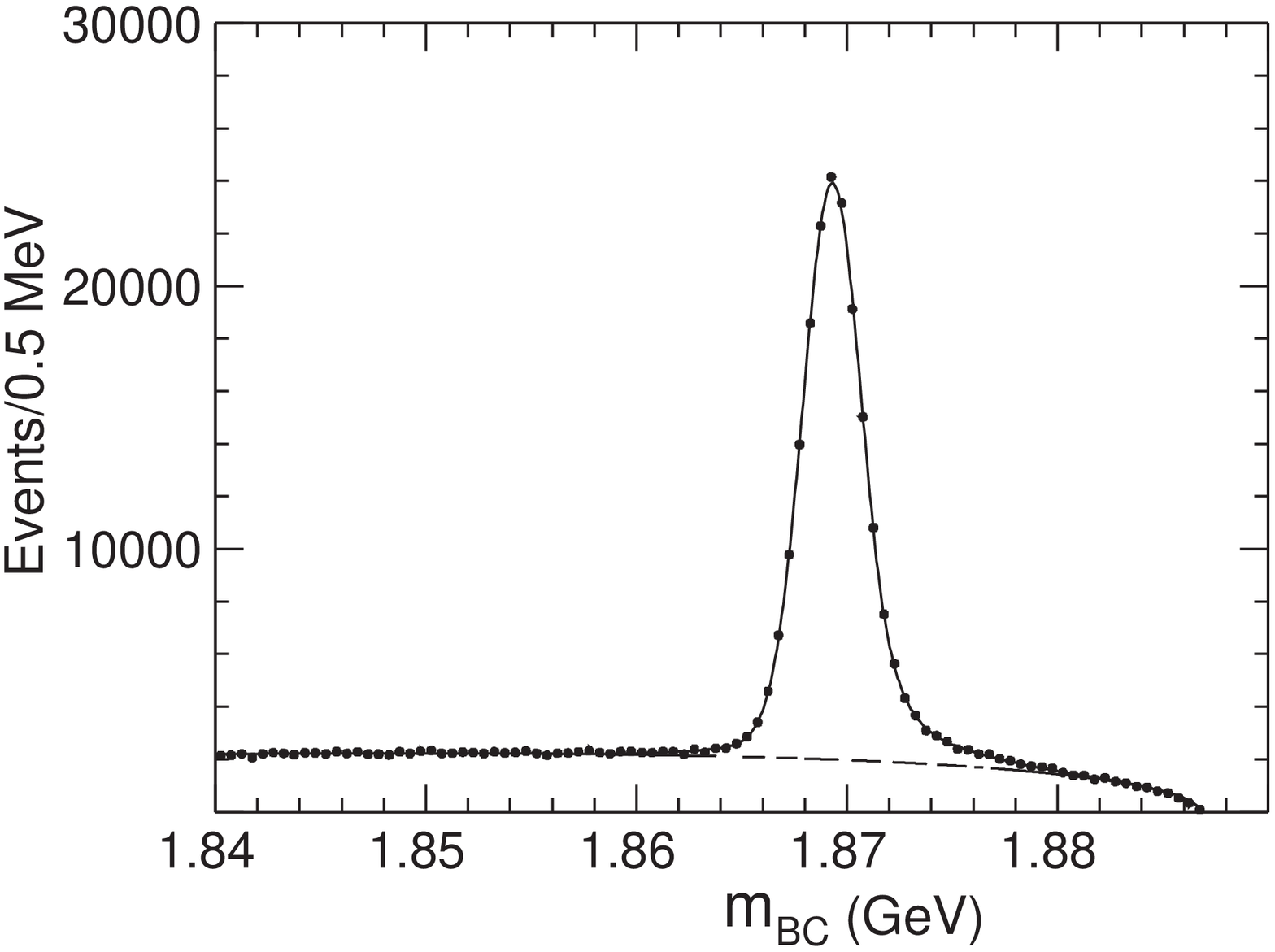}
       \includegraphics[width=0.50\textwidth]{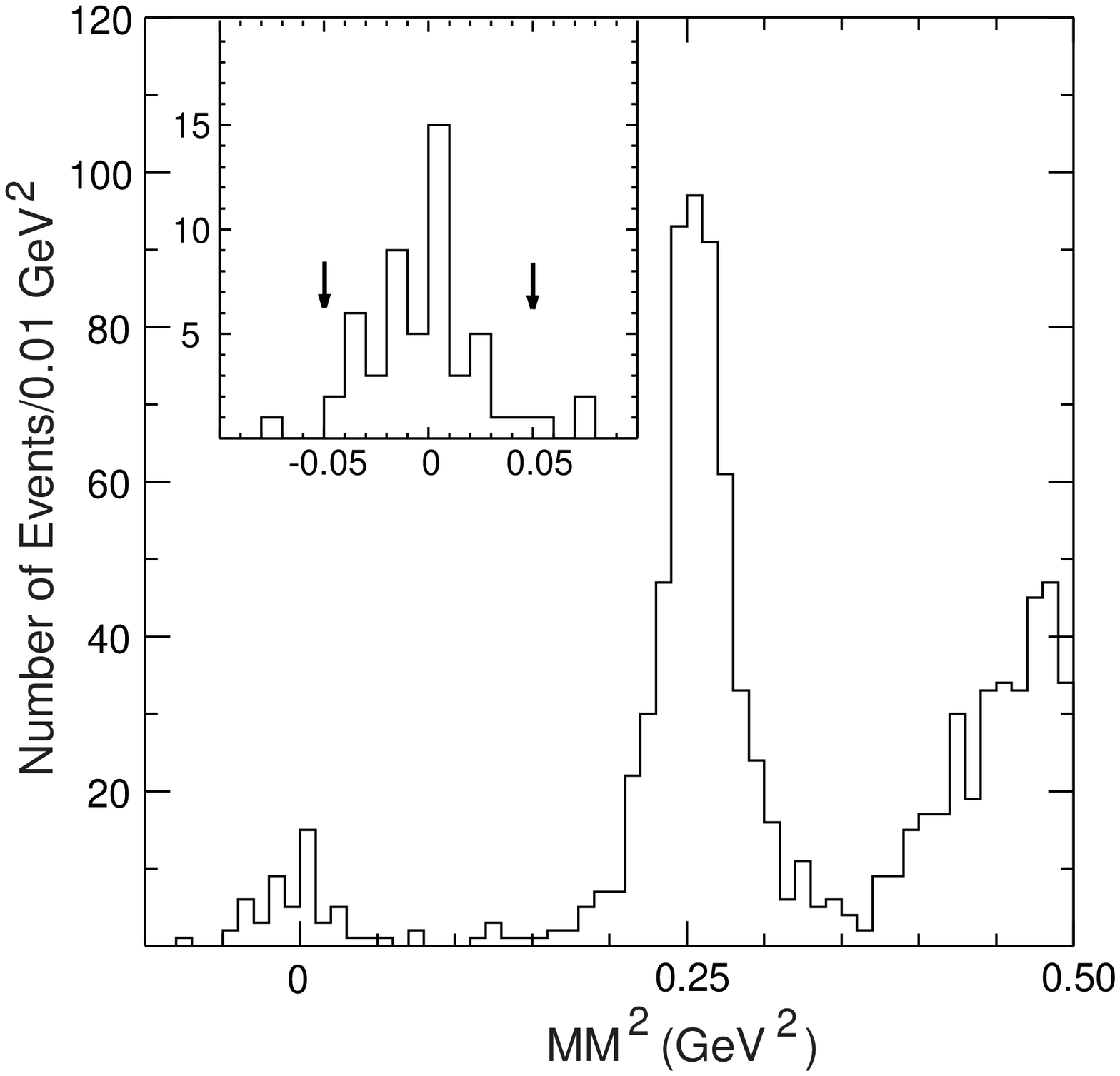}}}
\caption{Left: Beam-constrained mass for the sum of fully reconstructed
         $D^-$ decay candidates. The solid curve shows the fit to the sum
         of signal and background functions, while the dashed curve
         indicates the background.
         Right: MM$^2$ using $D^-$ tags and one additional opposite sign
         charged track and no extra energetic clusters (see text). 
         The insert shows the signal region for $D^+\to\mu^+\nu$ 
         enlarged; the defined signal region is shown between the two 
         arrows.
         } 
\label{Dreconnew}
\end{figure}

Using our sample of $D^-$ candidates we search for events with a
single additional charged track presumed to be a $\mu^+$. The track
must make an angle $>$35.9$^{\circ}$ with respect to the beam-line,
deposit less than 300 MeV of energy in the calorimeter,
characteristic of a minimum ionizing particle, and not be
identified as a kaon.
 Then
we infer the existence of the neutrino by requiring a measured value
near zero (the $\nu$ mass squared) of the missing mass squared
defined as
\begin{equation}
{\rm MM}^2=\left(E_{\rm
beam}-E_{\mu^+}\right)^2-\left(-\textit{\textbf{p}}_{D^-}
-\textit{\textbf{p}}_{\mu^+}\right)^2, \label{eq:MMsq}
\end{equation}
where $\textit{\textbf{p}}_{D^-}$ is the three-momentum of the fully
reconstructed $D^-$.

In order to restrict the sample to candidate $\mu^+ \nu$ events, we
select events with only one charged track in addition to the tagging
$D^-$. Events with extra tracks originating within 0.5 m (radially)
of the event vertex are rejected, as are events having a maximum
neutral energy cluster of more than 250 MeV. These cuts are highly
effective in reducing backgrounds especially from $D^+\to
\pi^+\pi^0$ decays, but they introduce an inefficiency because the
decay products of the tagging $D^-$ can interact in the detector
material leaving spurious tracks or clusters.

\section{Results}

The MM$^2$ distribution is shown in Fig.~\ref{Dreconnew}.  We see a peak
near zero containing 50 events within the interval $-0.050$ GeV$^2$
to +0.050 GeV$^2$, approximately $\pm 2\sigma$ wide. The peak is
mostly due to $D^+\to\mu^+\nu$ signal. The large peak centered near
0.25 GeV$^2$ is from the decay $D^+\to \overline{K}^0\pi^+$ that is
far from our signal region and is expected, since many $K_L$ escape
our detector.

There are several potential background sources; these include other
$D^+$ modes, misidentified $D^0\overline{D}^0$ events, and continuum
including $e^+e^-\to\gamma\psi'$. Hadronic sources need to be
considered because the requirement of the muon depositing less than
300 MeV in the calorimeter, while about 99\% efficient on muons,
rejects only about 40\% of pions or kaons as determined from a pure
sample of $D^0\to K^{-}\pi^{+}$ decays.

There are a few specific $D^+$ decay modes that contribute
unwanted events in the signal region. Residual $\pi^+\pi^0$
background is determined from a simulation that uses a branching
fraction of (0.13$\pm$0.02)\% \cite{CLEOpipi} and yields
1.40$\pm$0.18$\pm$0.22 events; the first error is due to Monte
Carlo statistics, and the second is systematic, due mostly to the
branching ratio uncertainty. We find background from
$D^+\to\tau^+\nu$ only when $\tau^+\to \pi^+\nu$. Since the
$\tau^+\nu$ branching ratio is known to be 2.65 times the
$\mu^+\nu$ rate from Eq.~\ref{eq:equ_rate}, our simulation gives
1.08$\pm$0.15$\pm$0.16 events, where the systematic error arises
from our final uncertainty on the $\mu^+\nu$ decay rate. The
$\overline{K}^o\pi^+$ mode (branching ratio of (2.77$\pm$0.18)\%
\cite{PDG}) gives a large peak in the MM$^2$ spectrum near 0.25
GeV$^2$. While far from our signal region, the tail of the
distribution can contribute. 
Our total background is 2.81$\pm$0.30$\pm$0.27 events. 
We have 47.2$\pm 7.1^{+0.3}_{-0.8}$ $\mu^+\nu$ signal events after
subtracting background. The detection efficiency for the single muon
of 69.4\% includes the selection on MM$^2$ within $\pm2\sigma$
limits, the tracking, the particle identification, probability of
the crystal energy being less than 300 MeV, and corrections for
final state radiation. It does not include the
96.1\% efficiency of not having another unmatched cluster in the
event with energy greater than 250 MeV. We also need to account for
the fact that it is easier to find tags in $\mu^+\nu$ events than in
generic decays by a small amount, (1.5$\pm$0.4$\pm$0.5)\%, as
determined by Monte Carlo simulation.

Our result for the branching fraction, using the tag sum in
Table~\ref{tab:Dreconnew}, is
\begin{equation}
{\cal{B}}(D^+\to\mu^+\nu)=(4.40\pm 0.66^{+0.09}_{-0.12})\times
10^{-4}~.
\end{equation}

The decay constant $f_{D^+}$ is then obtained from
Eq.~(\ref{eq:equ_rate}) using 1.040$\pm$0.007 ps as the $D^+$
lifetime \cite{PDG}, and $|V_{cd}|$ = 0.2238$\pm$0.0029
\cite{KTeV}. (We add these two small additional sources of
uncertainty into the systematic error.) Our final result is
\begin{equation}
f_{D^+}=(222.6\pm 16.7^{+2.8}_{-3.4})~{\rm MeV}~.
\end{equation}

We use the same tag sample to search for $D^+\to e^+\nu_{e}$. We
identify the electron using a match between the momentum measurement
in the tracking system and the energy deposited in the CsI
calorimeter as well as insuring that the shape of the energy
distribution among the crystals is consistent with that expected for
an electromagnetic shower. Other cuts remain the same. We do not
find any candidates, yielding a 90\% C.L. limit of ${\cal{B}}(D^+\to
e^+\nu_{e})<2.4\times 10^{-5}~.$

\end{document}